# Effective Road Model for Congestion Control in VANETs

Dongre Manoj M.1 Bawane Narendra G. 2

[1]Department Electronics and Telecommunication Engg. Ramrao Adik Institute of Technology ,NaviMumbai, India
mmdongre04@gmail.com
2Department of Computer Engg. S.B.Jain Institute of Management & Technology , Nagpur ,India
narendra.bawane@yahoo.com

## ABSTRACT

*Congestion on the roads is a key problem to deal with, which wastes valuable time.. Due to high mobility rate and relative speed link failure occur very often. VANET is used to tackle the problem of congestion, and make decisions well in advance to avoid traffic congestion.  In this paper we proposed a solution to detect and control the traffic congestion by using of both (V2V) and (V2I), as a result the drivers become aware of the location of congestion as well as way to avoid getting stuck in congestion. The congestion is detected by analyzing the data obtained by vehicular communication and road side units to avoid the traffic. Our proposition system is competent of detecting and controlling traffic congestion in real-time. V2V and V2I communication network is used to receive and send the messages. We  simulate the result by using Congestion Detection and Control Algorithm (CDCA), and show that this is one effective way to control congestion. The Proposed methodology ensures reliable and timely delivery of messages to know about congestion and avoid it.*

## 1 INTRODUCTION

Vehicular Ad-hoc network (VANET) is a possible solution to design networks that can solve traffic congestion problems [1,2],V2V is used for short range communication(1Km).Vehicle equipped with  V2V devices  can serve the purpose of dedicated short range communication or DSRC[3,4].. If density of vehicles is low or if the vehicle is breaks down, V2V will cease to be effective, it cannot solve the problems of early warning over a large range so as to allow vehicles to take alternate routes [5, 6, 7].The V2I is the other possibility which offers larger range and other benefits that occur from longer range. It has long range and data processing and transmission of message to a full region covering all vehicles, but large resource is needed as communication towers are very expensive [8]. V2I Communication range depends on its power and spacing between the towers [9, 10, and 11.] Traffic congestion is a serious problem on express Highway and in urban areas. The accident happens due to the misbehavior of the driver, weather conditions and failure of vehicles. Due to this, vehicles are either stand still or moving with very low speed. Few hours or few days it may take to clear the Traffic Congestion.

Thus we propose efficient and intelligent solution.

a) Identify and notice the congestion, its place (b) Warning message broadcast continuously by Affected vehicles to the nearby vehicles by V2V Communication and for long range V2I Communication (c) After receiving the message the follower vehicle takes decision,

To avoid heavy traffic and to control congestion we propose the (CDCA) Algorithm Our aim is to find a solution which enables the vehicles to take intelligent decisions based on the latest traffic status updates by broadcasting information among the vehicles gives prior information of the road condition and helps vehicles to take prior decision to prevent traffic congestion.

An overview of related work explained in section- 2, the methodology in section -3. Section- 4 covers algorithm and system flow charts. Model validation and analysis explain in Section-5 and -6. Finally section -7.concluded the paper.

## 2. RELATED WORK

Over wireless network many schemes and algorithms were studied, experimented to solve traffic Congestion problems in ref [1] discuss about Congestion Detection using CDCA algorithm and shows the one effective way to detect congestion. In Ref [2] multi agent based congestion control scheme are explained, which intensely get hold of the status of the adjacent node information and decides the safe delivery of packets with congestion free path. This gives guarantees the trustworthiness and messages delivery. In Ref [3] implement an algorithm for fine-tuning the Distributed Fair transmits Power Adjustment for VANETs congestion control approach. The beacon congestion matter in VANET were deal in ref [4] here congestion detection and control scheme were explain ,safety application have need of timely and steadfast broadcasting of the event driven forewarning message address.

In ref [5] proposed and implement algorithms here carrier sense (CS) threshold value can be assigned vigorously for fine tuning, the distributed fair transmit power adjustment D-FTPA of VANET congestion control approach. The D-FTPA algorithm can be used any situation i.e. traffic and non traffic condition. The study of exiting congestion control algorithm to solve the congestion problems were discuss in ref.[6] also depict the limitation and advantages of congestion control algorithm. Adaptive multichannel approach discuss in ref.[7] based on current traffic condition allow the flexible multichannel usage , average packet delay and throughput parameters are studied for system performance. Probability model checking techniques were used to analyzed the congestion control protocols also investigate the correctness of the system were explain in ref [8].

In ref[9] discuss the research challenges and issues also the explain the various techniques which control the load on radio channel, cooperative system were explain for congestion control. Using different system parameter the routing protocol performance were discuss and simulate using NS-2 Network simulator in ref[10]. MAC-to-MAC delay reliability, Decentralized congestion control (DCC) state stability parameters were consider for performance analysis to control the congestion in VANET , also active DCC multistate method Proposed in ref.[13]

## 3. METHODOLOGY

The communication is carried out by the affected vehicle. The warning message continuously broadcast by affected vehicle. Both (V2V) and (V2I) communication are considered for vehicle in short range and long range respectively   the different system parameters are explained below.

*Data Collection:* The most basic thing is to collect data from the environment. Current location and speed.

*Sharing of Information:* The information must be shared between vehicles. Using V2V and V2I Comunication.

*Message Broadcast:* The warning message broadcast continuously by affected vehicle, the vehicle in the range can receive the message and take decision accordingly.

*Decision Making:* Appropriate decision has to be taken so that the congestion can be detected, and accordingly the vehicle in the range takes decision after updating message field.

*Bandwidth:* For congestion control needs to make intelligent use of the bandwidth. Bandwidth utilization is very important for sending large amounts of information, otherwise will overload the system.

*Vehicles Requirements:* All Vehicles should be equipped with the same hardware which allows sharing and collection of information's. A Novel system has to be designed so that it can work even all vehicles on the road participate in the congestion detection system.

### 3.1 Congestion Detection and Control Algorithm (CDCA)

In the preparation of algorithm Possibility of an event of an accident condition is considered. The affected vehicle broadcast warning message continuously i.e. alert message regarding accident or about worst road condition. The vehicle in the short range and long range received the warning message and alert by using V2V & V2I communication respectively

 Steps in CDCA algorithm are as follows.

a) The affected vehicle of an accident event occur only broadcast the information.
b)  Then affected vehicle broadcast the message (information) to the nearby vehicle

c) The vehicle which is in the short distance range the messages are transmitted by the V2V Communication whereas vehicle in long distance the messages are transmitted by the road side units. (V2I) Communication.

d) Each vehicle in the communication must have a unique Identification number (ID).

e) Speed of the vehicle.

f) The status of the lane to show that particular lane is blocked.

g) After receiving the message the nearby vehicle will take proper decision for diversion and then updates the message field and forwards it to other vehicles.

h) Next incoming vehicle gets decision about the blocked lane and divert form it.

i) The roadside units are also getting warning message that accident happened in particular lane, according they starts broadcasting message which assists in congestion control. This is long range communication so well in advance vehicles are getting message of slow down the speed, they have sufficient time to take proper decision and helps in avoiding the congestion.

j) Due to short range and long range communication the message are received by all the vehicles coming in range.

k) The lane changing vehicle according to the decision immediately stops the broadcasting the message thus limiting the number of messages getting broadcasted and also limiting the overloading of the system.

## 4. Flowcharts for V2V & V2I Communication:

The different flowcharts of message broadcasted by affected vehicle, of message broadcasted by short range vehicle and of Message Broadcasting by Tower are explained in below.

## 4.1. V2V Communication:

The speed of vehicle becomes zero as accident event occur. The warning message broadcast continuously by the affected vehicle and keep on checking for vehicle in range. As soon as vehicle in the range i.e. (V2V) communication, affected vehicle transmit the message and will be broadcast continuously for other vehicles in the lane. The flowchart of accident event shown in fig.1

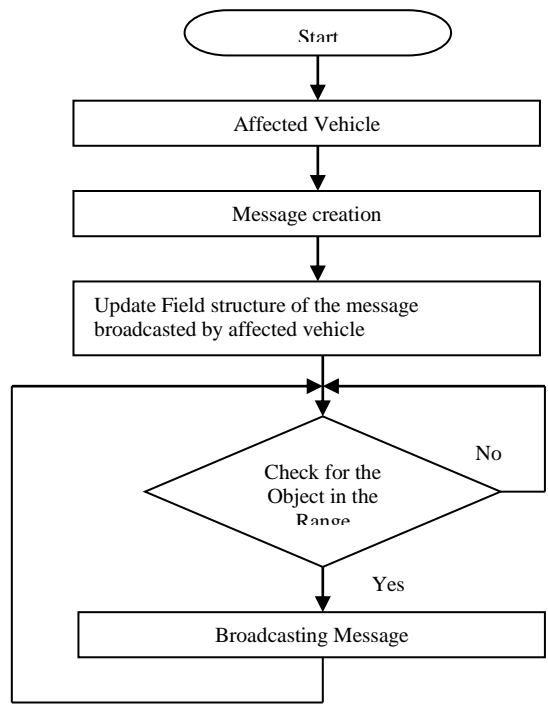

Fig.1. Flowchart for accidents event.

The vehicle in the range receives the warning message transmitted by affected vehicle, vehicle gets the information about the status of lane and situation and the vehicle takes decision of diversion, also updates the message fields and retransmits the message. The receiving vehicle check the current lane of vehicle is equal to the blocked lane. Flowchart is shown in fig. 2

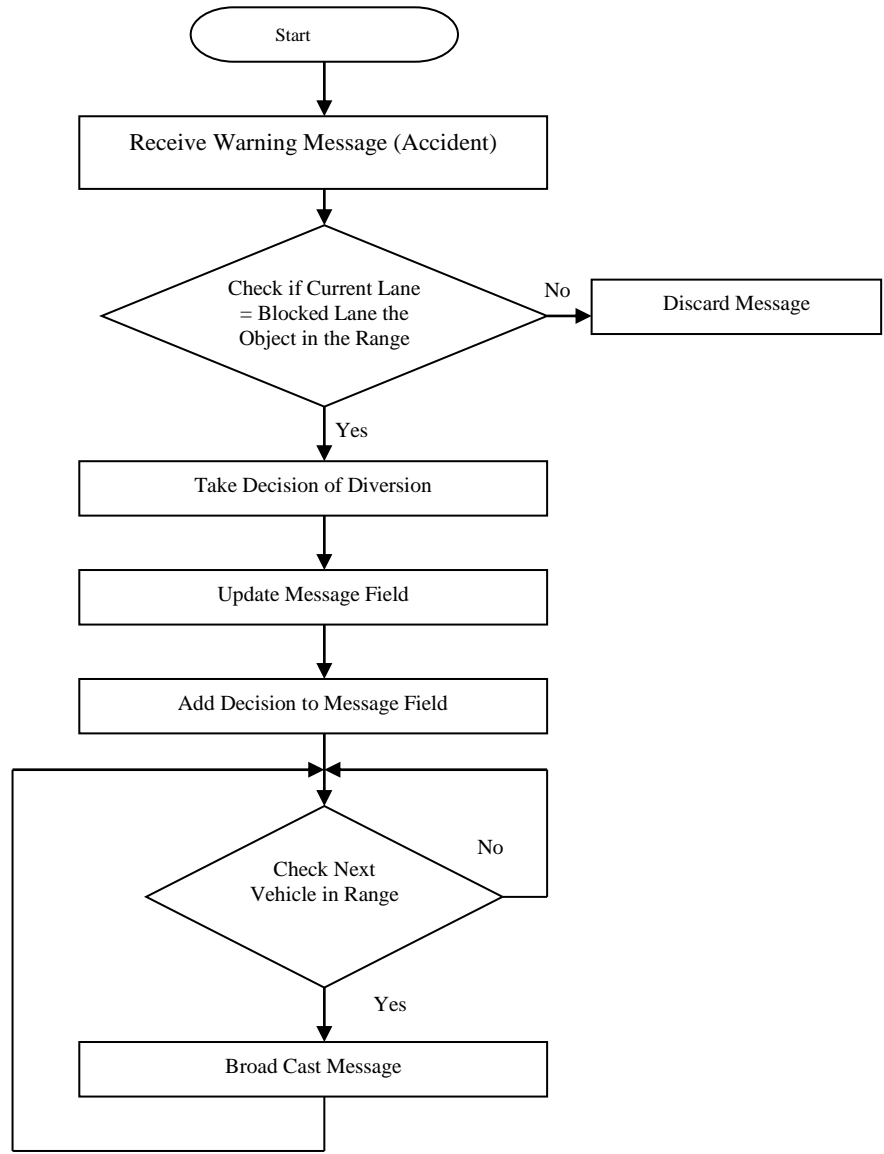

Fig.2. Flowchart for V2V communication .

## 4.1. V2I Communication:

When the vehicle is in the long range then the Message is received from the road side unit. i.e. tower is broadcast the message which are received by affected vehicle. The vehicles in long range received message from tower and take the suitable decision. The flowchart is as shown in fig.3.

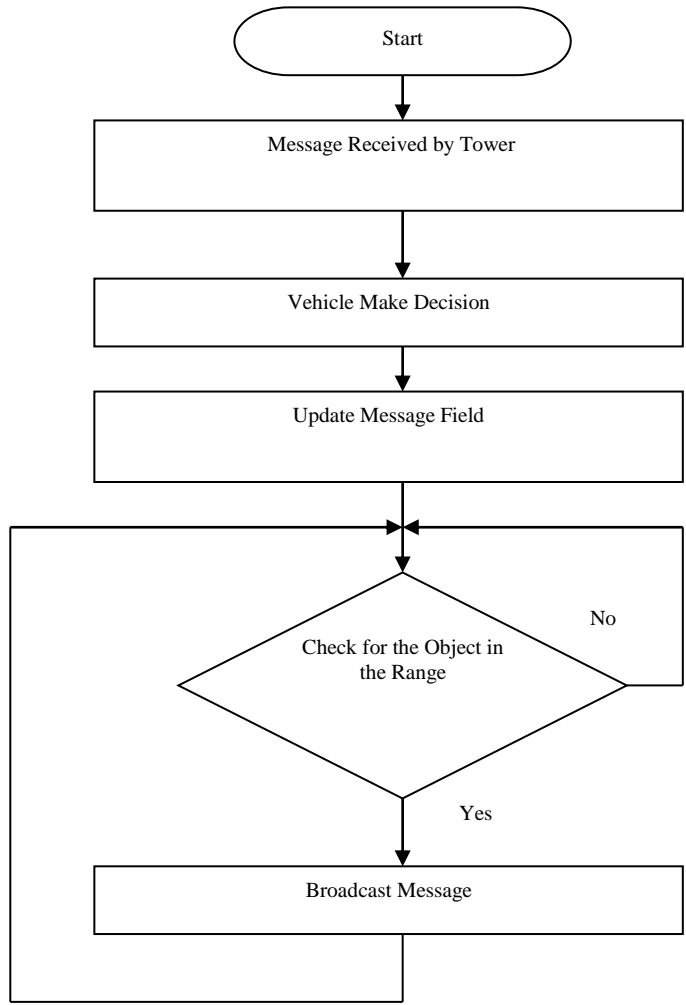

Fig. 3 Flowchart for V2I communication.

## 5. MODEL VALIDATION:

To validate the road model simulation has been developed in Net Beans IDE 7.0 platform. The operating system window 8 and communication pattern specially Geobroadcast pattern are used for visualization and analysis .The basic Configurations parameters and description are as shown in table no 1.

TABLE No. 1: Basic Configuration Parameters

| Parameters | Values |
| --- | --- |
| Platform | Net Beans IDE 7.0 |
| Operating system | Windows 8 |
| Broadcast pattern | Geobroadcast |
| Simulator | Traffic simulator |
| Vehicle type | Truck & car |
| Speed of vehicle | Car :108 km/h<br>Truck :54km/h |
| No of vehicle on road | 500 |
| Road type | Main flow &<br>Ramp flow |
| No of lanes in each direction | 3 in each |
| Vehicle percentage | 20% |
| Changing threshold | 0.2 |
| Simulation speed | 10.0 |
| Impose speed limit | 80km/h |
| Politeness factor | 0.25 |

## 5.1. Vehicles on Road

The system model scenario as shown in fig.4. In this platform we shows vehicle running through all three lanes.

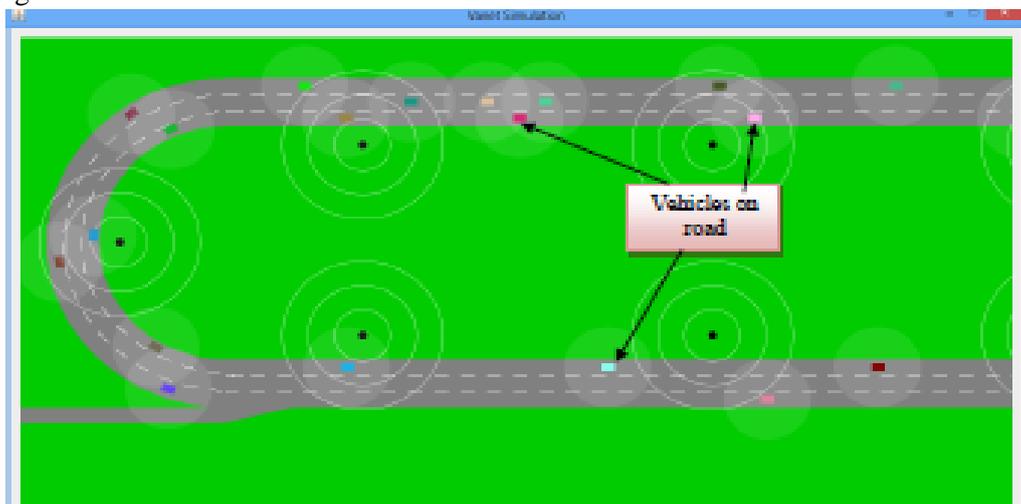

Fig. 4. Vehicles on road

## 5.2. Congestion Scenes

We create a Congestion scenario by stopping the vehicle in both upper and lower lane, middle lane will be free. The scenario is shown in the fig. 5

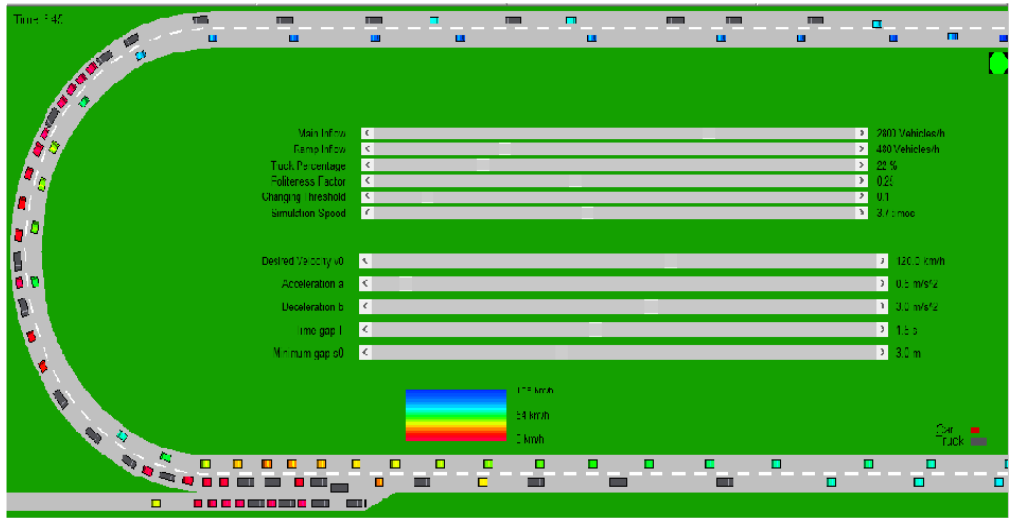

Fig. 5. Congestion scenario

Figure no 6 shows the accidental condition in which the vehicle is blocked in lane number and other vehicles without creating congestion are taking diversions.

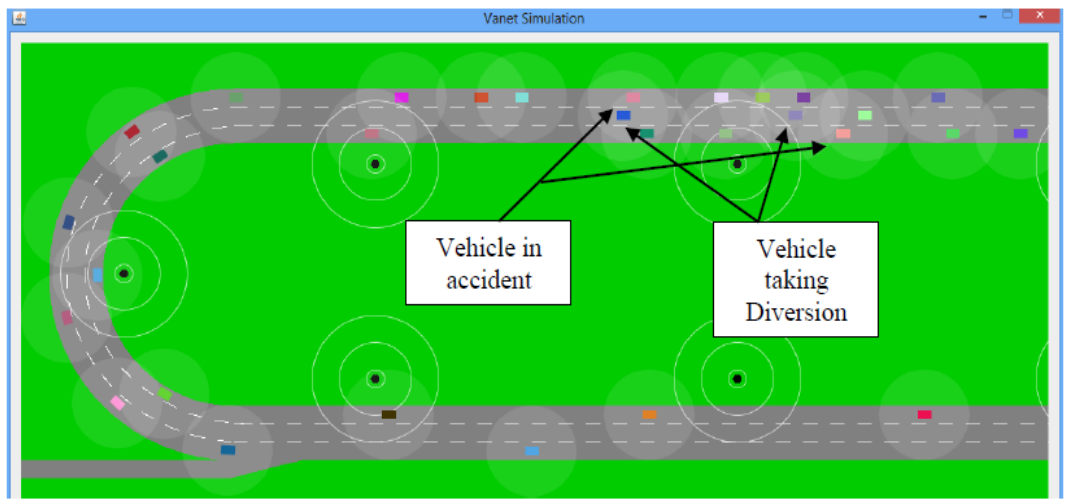

Fig. 6. Accident Events in upper and lower lane

# 6. ANALYSIS

## 6.1. Congestion Detection

The vehicle in Lane 1 & lane 2 is in congestion due to zero speed, and no vehicle in congestion in lane 3. The graph is plotted between the numbers of vehicles v/s speed and by considering different time series find out how many vehicles are in congestion at particular time as shown in fig. 7.

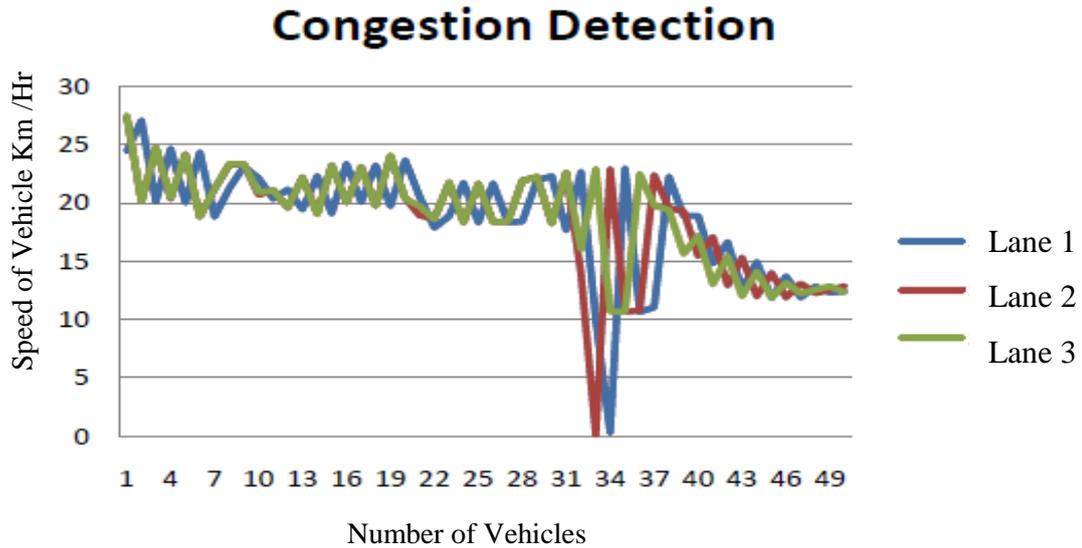

Fig. 7. Congestion Detection

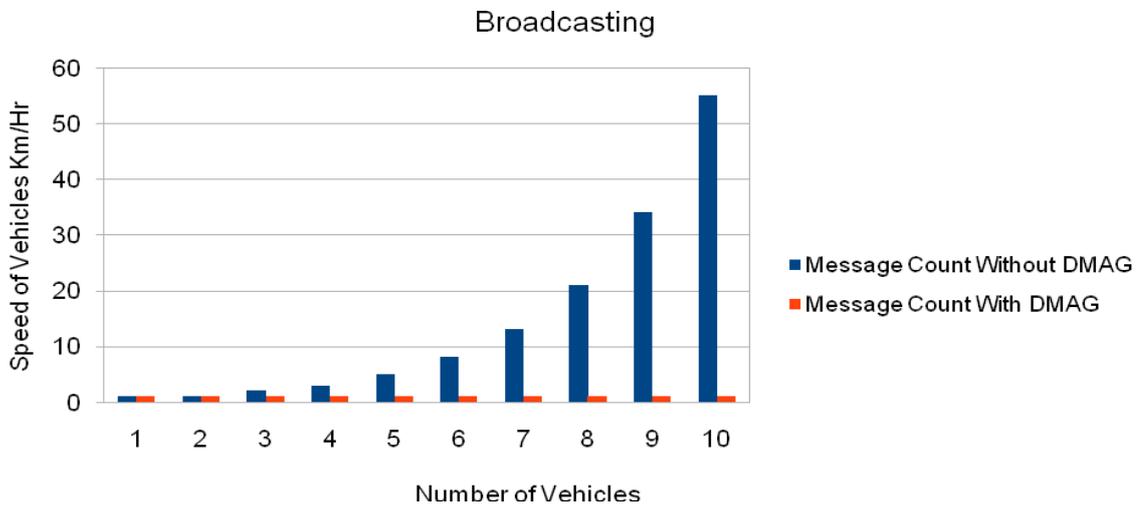

Fig.8. Graph of Congestion detection

From graph we observe that the speed of vehicle become zero with algorithm hence the congestion is detected. Without algorithm no vehicles are in congestion as the speed is not reducing to zero. Due to single message broadcast by the Vehicle the network overhead is reduced hence improving the bandwidth utilization.

## 6.2 Congestion Control

We analyse that there is no vehicle in congestion which indicates the effectiveness of the algorithm and thus our model is effective which can control the congestion as shown in fig.9

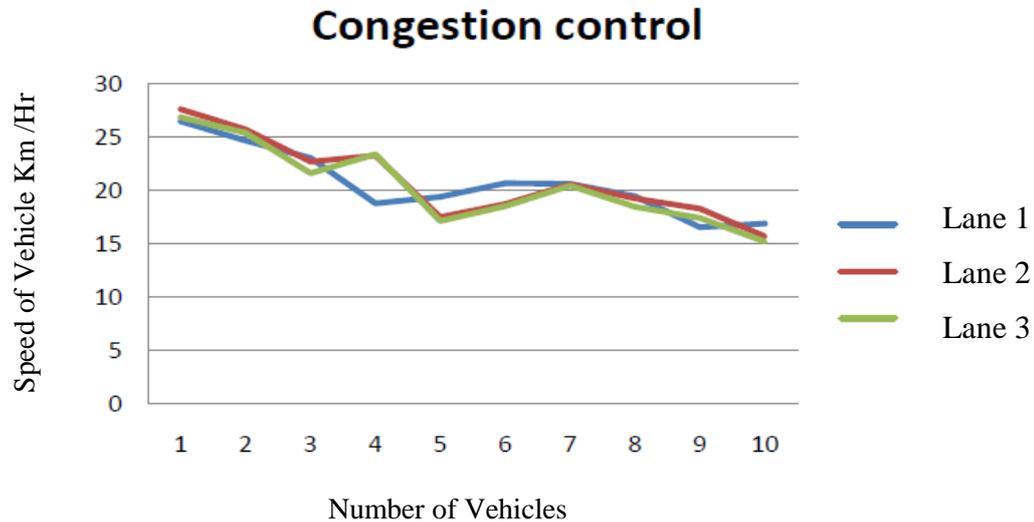

Fig. 9   Congestion Control.

## 7. CONCLUSION

In this paper we propose an effective road model for congestion control based on VANET. The status, different events happening on road and its condition information is broadcasted to alert the drivers about the congestion ahead. This timely information is useful for taking decision as per the broadcasted message. The speed of the vehicle reducing to zero as the Algorithm is applied hence we control and  detect the traffic congestion,  only one message is broadcasted by the  vehicle which reduces the overhead on the network improving bandwidth utilization, which shows the effectiveness of the system hence it is   an intelligent system to detect and control the traffic congestion.  We plan to extend this research work to control Inner-city traffic and address security issues to improve the intelligent transportation system (ITS) which is the need of the day.

Authers:

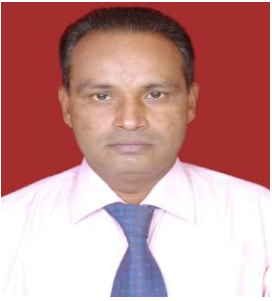

**Manoj M. Dongre,** received the B.E (Electronics Engineering), from Dr. Babasaheb Ambedkar Marathwada University, Aurangabad (MS) and M.Tech (Electronics and telecommunication Engineering) from Dr. Babasaheb Ambedkar Technical University, Lonere, Raigad (MS) in 1995 and 2007 respectively. Currently he is pursuing PhD. in the department of Electronics Engineering at Nagpur University. He is currently Assistant Professor with the department of Electronics and Telecommunication Engineering in Ramrao Adik Institute of Technology, Mumbai University, India. He has having an overall teaching experience of 19 years. His research interest includes wireless network, VANET, Mobile Commuication and Optical Network.

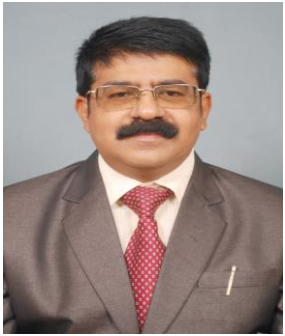

**Dr. Narendra G. Bawane** received the B.E. from Nagpur University in 1987 and M. Tech. in 1992 from IIT, New Delhi. He has completed his Ph. D. from VNIT in 2006; currently he is working as Professor and Principal of S. B. JNIT, Nagpur since 2011. He is having total teaching experience of more than 25 years to his credit including 12 years in Govt. organization. He is an IEEE senior member and supporting IEEE Bombay section which covers Maharashtra, MP, Chhattisgarh & Goa as SMC Chair and member executive committee form 2014. He is life member of professional society like CSI and ISTE. He has an author of more than 74 research articlesand papers in various international journals and conferences. He has authored two books. His area of interest includes soft computing, image processing, BCI and networking. He is handling several research projects and strongly believes in innovation and creativity.

.